\documentclass[a4paper,12pt]{article}
\usepackage{lmodern}    
\usepackage{graphicx}   
\usepackage{amsmath}    
\usepackage{amsthm}     
\usepackage{amssymb}    
\usepackage{abstract}
\usepackage{cite}      
\usepackage{color}      
\usepackage{ulem}       
\usepackage{enumerate}  
\usepackage{enumitem}   
\usepackage{tikz}       
\usepackage{mathrsfs}
\usetikzlibrary{arrows.meta} 
\usepackage{caption}    
\usepackage[top=2.5cm,bottom=2.3cm,left=2.5cm,right=2.2cm]{geometry}
\usepackage[colorlinks,linkcolor=blue,anchorcolor=blue,citecolor=blue]{hyperref} 
\numberwithin{equation}{section} 
\usepackage[numbers,sort&compress]{natbib} 
\usepackage{booktabs}   
\usepackage{longtable}  
\usepackage{threeparttable} 
\allowdisplaybreaks[3]  
\usepackage{authblk}    
\newtheorem{theorem}{Theorem}[section]

\newtheorem{proposition}[theorem]{Proposition}
\theoremstyle{definition}

\newtheorem{remark}[theorem]{Remark}

\usepackage{xcolor}

\usepackage{ragged2e}
\justifying\let\raggedright\justifying 

\title{The non-Abelian two-dimensional Toda lattice and matrix sine-Gordon equations with self-consistent sources}
\author[1]{Mengyuan Cui}
\author[1,\footnote{5572@cnu.edu.cn}]{Chunxia Li}
\affil[1]{School of Mathematical Sciences, Capital Normal University, Beijing 100048, China}
\date{}

\begin{document}

\maketitle

 \begin{abstract}
The non-Abelian two-dimensional Toda lattice and matrix sine-Gordon equations with self-consistent sources are established and solved. Two families of quasideterminant solutions are presented for the non-Abelian two-dimensional Toda lattice with  self-consistent
sources. By employing periodic and quasi-periodic reductions, a matrix sine-Gordon equation with self-consistent sources is constructed for the first time, for which exact solutions in terms of quasideterminants are derived.  
\end{abstract}



\section{Introduction}
As one of the main organizing tools in noncommutative (NC) algebra, quasideterminants were first introduced by
Gelfand and Retakh in \cite{gelfand2005quasideterminants} and have been found applications in many areas including NC symmetric functions, NC integrable systems, quantum algebra and Yangians. Quasideterminants play crucial roles in the NC soliton theory. They can be used to treat NC integrable systems in a unified way without regard to the reasons for noncommutativity and greatly simplify proofs in commutative theories as well. In particular, NC extensions of integrable systems have received much attention and various integrability has been revealed \cite{paniak2001exact,carillo2009noncommutative,hamanaka2003noncommutative,dimakis2004extension,hamanaka2003towards,hamanaka2005commuting,gilson2007direct,wang2003exact,gilson2008direct,wang2004noncommutative,mulase1988solvability,adler1998random,gilson2020soliton}.

As one of the most notable examples, the non-Abelian two-dimensional Toda lattice ($2$DTL) was first studied in \cite{mikhaǐlov1979integrability} with its Darboux transformation given in \cite{sall1982darboux}. Later on, the following generalized non-Abelian $2$DTL was proposed 
\begin{align}
&U_{n,x}+U_nV_{n+1}-V_nU_n=0,\label{nncTL1}\\
&V_{n,t}+\alpha_nU_{n-1}-U_n\alpha_{n+1}=0\label{nncTL2}
\end{align}
and Darboux and binary Darboux transformations were obtained \cite{nimmo1997darboux}. Under the transformations
\begin{equation*}
U_n=X_nX_{n+1}^{-1},\quad V_n=X_{n,x}X_n^{-1},\label{nT}
\end{equation*}
the generalized non-Abelian $2$DTL  \eqref{nncTL1} and \eqref{nncTL2} transform into 
\begin{align}\label{nc2dTL}
    (X_{n,x}X_n^{-1})_t+\alpha_nX_{n-1}X_n^{-1}-X_nX_{n+1}^{-1}\alpha_{n+1}=0.
\end{align}
In \cite{li2008quasideterminant}, both quasiwronskian and quasigrammian solutions were constructed for \eqref{nc2dTL} by Darboux and binary Darboux transformations. Consequently, a matrix sine-Gordon (sG) equation together with its exact solutions in term of quasideterminants were derived as the $2$-periodic reductions.     


Soliton equations with self-consistent sources (SESCSs) have important applications in physics. In general, the existence of sources can result in solitary waves moving with nonconstant velocity or cause a great
variety of dynamics of soliton solutions. In literature, some meaningful SESCSs haven been studied by different ways \cite{mel1989interaction,mel1989capture,leon1990solution,yunbo1996lax,zeng1998constructing,zeng1999bi}.
 As far as we are concerned, the $2$DTL ESCS as well as its integrability are derived in \cite{wang20072d} by source generation method.  
Later on, an extended $2$DTL hierarchy is obtained by its squared eigenfunction symmetries which includes the $2$DTL ESCS as its special case
\cite{liu2008extended}.
As is well known that the $2$-periodic reduction of the standard $2$DTL leads to the scalar sG equation.
It has been also shown that the $2$-periodic reduction of the $2$DTL ESCS leads to the sG equation with self-consistent sources \cite{liu2011generalized}. In this paper, we are going to explore how to establish the non-Abelian $2$DTL ESCS, what integrability the non-Abelian $2$DTL ESCS possesses and the corresponding periodic reductions.  

%


 The paper is organized as follows. The non-Abelian $2$DTL ESCS is established and solved in Section \ref{sec3}. A matrix sG ESCS and its exact solutions in terms of quasideterminant are derived by period reductions in Section \ref{sec4}. Section \ref{conclusion} is devoted to conclusions and discussions.
\section{ The non-Abelian $2$DTL ESCS} \label{sec3}
In this section, we will construct and solve the non-Abelian $2$DTL ESCS equation by source generation method based on the existing quasiwronskian and quasigrammian solutions of the  non-Abelian $2$DTL equation. Consequently, both quasiwronskian and quasigrammian solutions are presented for the non-Abelian $2$DTL ESCS. The connections of the non-Abelian $2$DTL ESCS with the nonlinear $2$DTL ESCS derived by two different ways are clarified as well, which implies the advantage of dealing with NC integrable systems over commutative integrable systems. As for the properties of quasideterminants, one can refer to \cite{gelfand2005quasideterminants,gilson2007direct} for details. 
\subsection{Quasiwronskian solutions }\label{sub2.1}
It has been proved that the  non-Abelian $2$DTL equation \eqref{nc2dTL} has the quasiwronskian solution \cite{li2008quasideterminant}
\begin{equation*}
\begin{aligned}
   {Q}_n=
      \left|\begin{matrix}
       \Theta_{n}           &       \boxed{0}\\
       \tilde{\Theta}_{n}   &        e_N  
       \end{matrix}\right|,                  
\end{aligned}
\end{equation*}
where $e_i$ is the column vector of length $N$ with $1$ in the $i$th row and $0$ elsewhere, and $\tilde{\Theta}_{n}=(\Theta_{n+1},\Theta_{n+2},\cdots,\Theta_{n+N})^{T}$, $\Theta_{n}=({\theta}_{1,n},\dots,{\theta}_{N,n})$
with ${\theta}_{i,n}~ (i=1,2,\dots,N)$ satisfying the dispersion relations
\begin{equation}\label{dispersion1}
     ({\theta}_{i,n})_x={\theta}_{i,n-1},\quad 
     ({\theta}_{i,n})_t={\theta}_{i,n+1}. 
\end{equation}  
It is obvious that
\begin{equation}\label{Theta}
     \Theta_{n,x}=\Theta_{n-1},\quad 
    \Theta_{n,t}=\Theta_{n+1}. 
\end{equation}  

According to the source generation procedure, we assume that 
$${\theta}_{i,n}=f_{i,n}+(-1)^{i-1}g_{i,n}\large{C
}_{i}(t),$$
where $f_{i,n}$ and $g_{i,n}$ satisfy the dispersion relations (\ref{dispersion1}) and
$$C_i(t)=
\begin{cases}
    c_i(t),\quad\quad 1\le i\le K\le N,K\in Z^+,\\
    C_i, \quad\quad\quad  otherwise. 
\end{cases}$$
Then we have
\begin{equation*}
  ( {\theta}_{i,n})_x={\theta}_{i,n-1},\quad    ( {\theta}_{i,n})_t={\theta}_{i,n+1}+(-1)^{i-1}g_{i,n}C_{i}'(t), 
\end{equation*}
where $C_{i}'(t)$ denotes the $t$-derivative of $C_{i}(t)$. 

Denote 
$h_{i,n}=(-1)^{i-1}g_{i,n}$ and $H_{n}=(h_{1,n}c_{1}'(t),h_{2,n}c_{2}'(t),\cdots,h_{K,n}c_{K}'(t),0,\cdots,0)$, it follows that
\begin{equation*}\label{dispersion2}
    {\Theta}_{n,x}={\Theta}_{n-1},\quad
     {\Theta}_{n,t}={\Theta}_{n+1}+H_{n}.
\end{equation*} 
It is obvious that $X_n=Q_n$ no longer satisfies the  non-Abelian $2$DTL equation. As the expressions for other derivatives of ${Q}_n$ remain the same as in the case of the  non-Abelian $2$DTL equation
 \eqref{nc2dTL}, it is sufficient to compute $({Q}_{n,x}Q_n^{-1})_t$. Notice that
\begin{equation*}  
    {Q}_{n,x}=\left(
       \left|\begin{matrix}
       {  {\Theta}}_{n-1}      &       \boxed{0}\\
       \tilde{  {\Theta}}_n    &        e_N   
       \end{matrix}\right|
     + \left|\begin{matrix}
         {\Theta}_{n}      &       \boxed{0}\\
       \tilde{  {\Theta}}_n    &       e_1   
       \end{matrix}\right|
     \left|\begin{matrix}
         {\Theta}_{n}      &       \boxed{0}\\
       \tilde{  {\Theta}}_n    &       e_N   
       \end{matrix}\right|  
       \right),
       \end{equation*}
Then we have
\begin{align*}
    ({{Q}_{n,x}} {Q}_n^{-1})_t=&
     {Q}_n {Q}_{n+1}^{-1}- {Q}_{n-1} {Q}_{n}^{-1}
      +\left|\begin{matrix}
          H_n               &     \boxed{0}\\
          \tilde{  {\Theta}}_n  &     e_1
        \end{matrix}\right|
      +  \sum_{i=1}^N 
        \left|\begin{matrix}
            {\Theta}_n               &     \boxed{0}\\
          \tilde{  {\Theta}}_n       &     e_i
        \end{matrix}\right|
        \left|\begin{matrix}
           H_{n+i}              &     \boxed{0}\\
          \tilde{  {\Theta}}_n      &     e_1
        \end{matrix}\right|\\
    &    -\left|\begin{matrix}
          H_{n-1}               &     \boxed{0}\\
          \tilde{  {\Theta}}_{n-1}  &     e_1
        \end{matrix}\right|
      -  \sum_{i=1}^{N-1} 
        \left|\begin{matrix}
            {\Theta}_{n-1}               &     \boxed{0}\\
          \tilde{  {\Theta}}_{n-1}       &     e_i
        \end{matrix}\right|
        \left|\begin{matrix}
           H_{n+i-1}                 &     \boxed{0}\\
          \tilde{  {\Theta}}_{n-1}       &     e_1
        \end{matrix}\right|.
\end{align*}
Further assume that $c'_i(t)=\beta_i(t)\eta_i(t)$ and introduce functions $q_{i,n}$ and $r_{i,n}$ as follows
\begin{equation*}\label{wronskian solutions}
 \begin{aligned}
 {q}_{i,n}=\left|\begin{matrix}
         {\Theta}_{n}          &    \boxed{h_{i,n}}\\
         {\tilde{\Theta}}_{n}  &        G_{i,n}   
\end{matrix}\right|\beta_i(t),\quad
       {r}_{i,n}=\eta_i(t)\left|\begin{matrix}
         e_i^{T}     &       \boxed{0} \\
          {\tilde{\Theta}}_{n-1}  &        e_1  
          \end{matrix}\right|,
     \end{aligned}
 \end{equation*}
 where $G_{i,n}=(h_{i,n+1},h_{i,n+2},\cdots,h_{i,n+K},0,\cdots,0)^{T}$.
Then we can derive the non-Abelian $2$DTL ESCS, which is stated in the following proposition. One can refer to Appendix \ref{sec:appendix-A} for the proof of the proposition.
\begin{proposition}\label{proposition-wron}
    The  non-Abelian $2$DTL ESCS takes the following form
\begin{align}
    &(X_{n,x}X_n^{-1})_t+X_{n-1}X_n^{-1}-X_nX_{n+1}^{-1}=\sum_{i=1}^K(q_{i,n}
    r_{i,n+1}-q_{i,n-1}r_{i,n}),\label{2dTLSCS}\\
    &q_{i,n,x}=X_{n,x}X_n^{-1}q_{i,n}+q_{i,n-1},\qquad \qquad \quad i=1,2,\cdots, K ,\label{2dTLSCS1}\\
   & r_{i,n,x}=-r_{i,n}X_{n,x}X_n^{-1}- r_{i,n+1},\qquad \qquad i=1,2,\cdots, K. \label{2dTLSCS2} 
\end{align}
\end{proposition}

\subsection{Quasigrammian solutions}\label{3.2}

The non-Abelian $2$DTL equation has been shown to have quasigrammian solutions expressed as \cite{li2008quasideterminant}
\begin{equation*}
   {R}_n=-\left|\begin{matrix}
             {\Omega}_n  &   P_n^{\dag}\\
             \Theta_n    &   \boxed{-I}  
             \end{matrix}\right|,
\end{equation*}
where $\Theta_n$ satisfies the same linear equations as in \eqref{Theta}, $P_n=(\rho_{1,n},\dots,\rho_{N,n})$ with $\rho_{i,n}$ satisfying the dispersion relations
\begin{equation*}\label{dispersion-rho}
   (\rho_{i,n})_x=-\rho_{i,n+1},\quad 
   (\rho_{i,n})_t=-\rho_{i,n-1}. 
\end{equation*}  
The potential ${\Omega}_n={\Omega}(\Theta_n,P_n)$ satisfies the following relations
\begin{align}\label{Omega-dispersion-1}
   \begin{split}
   {\Omega}_{n,x}=P_{n+1}^{\dag} \Theta_n,  \quad
   {\Omega}_{n,t}=-P_n^{\dag}\Theta_{n+1}, \quad
   {\Omega}_n-{\Omega}_{n-1}=-P_n^{\dag}\Theta_n,
   \end{split}
\end{align}
which lead to
$${\Omega}_n=C_{i,j}+\int{P_{n+1}^{{\dag}}\Theta_n}dx.$$

Next we assume that $C_{i,j}$ are no longer constants, but rather
$$C_{i,j}(t)=
\begin{cases}
    c_i(t),\quad\quad  i=j \quad and \quad 1\le i\le K\le N,\\
    C_{i,j}, \quad\qquad otherwise. 
\end{cases}$$
Then we can obtain
\begin{equation*}\label{Omega-dispersion-4}
       {\Omega}_{n,x}=P_{n+1}^{{\dag}}\Theta_n,  \quad
       {\Omega}_{n,t}=C'_{i,j}(t)- 
       P_n^{{\dag}}\Theta_{n+1},\quad  {\Omega}_n-{\Omega}_{n-1}=-P_n^{\dag}\Theta_n.
\end{equation*}
In this case, $X_n=R_n$  no longer satisfies the  non-Abelian $2$DTL equation. Since the
expressions for other derivatives of $R_n$ remain the same as in the case of the non-Abelian
$2$DTL equation \eqref{nc2dTL}, we only need to calculate $({R}_{n,x} {R}_n^{-1})_t$. Notice that 
\begin{equation*}
{R}_n^{-1}=\left|\begin{matrix}
         {\Omega}_{n-1}  &   P_n^{\dag}\\
            \Theta_n        &   \boxed{I}
    \end{matrix}\right|
    =I-\Theta_n  {\Omega}_{n-1}^{-1}P_n^{\dag},\end{equation*}
    Then we get
    \begin{equation*}
    ( {R}_{n,x} {R}_n^{-1})_t= {R}_{n-1} {R}_n^{-1}- {R}_n {R}_{n+1}^{-1}
    +\sum_{i=1}^K(\Theta_n {\Omega}_{n}^{-1}c'_i(t) {\Omega}_{n}^{-1}P_{n+1}^{\dag}-\Theta_{n-1} {\Omega}_{n-1}^{-1}c'_i(t) {\Omega}_{n-1}^{-1}P_{n}^{\dag}).
\end{equation*}
Introduce two sets of novel functions
\begin{equation*}\label{qr-gram}
   \begin{aligned}
    {q}_{i,n}=\left|\begin{matrix}
         {\Omega}_n   &     e_i^{T}  \\
        \Theta_n         &     \boxed{0}  
       \end{matrix}\right|\beta_i(t), \quad 
    {r}_{i,n}=\eta_i(t)
       \left|\begin{matrix}
         {\Omega}_{n-1}   &   P_n^{\dag}  \\
                e_i                     &    \boxed{0}  
       \end{matrix}\right|.
    \end{aligned}
\end{equation*}
 It can be proved that ${R}_n$, $ {q}_{i,n}$ and $ {r}_{i,n}$ form a set of quasigrammian solutions of the non-Abelian $2$DTL ESCS.  \eqref{2dTLSCS}-\eqref{2dTLSCS2}.
One can refer to Appendix \ref{appendixB} for detailed calculations.

 \subsection{From the non-Abelian $2$DTL ESCS to the  $2$DTL ESCS}\label{3.3}
Quasideterminants are closely related to determinants. In commutative setting, a quasideterminant is nothing but the ratio of two determinants. For this reason, the non-Abelian $2$DTL ESCS and the corresponding quasideterminant solutions can be used to yield the nonlinear $2$DTL ESCS and its determinant solutions. 


In commutative setting, the non-Abelian $2$DTL ESCS \eqref{2dTLSCS}-\eqref{2dTLSCS2} turns into
\begin{align}\label{cn-2dTL}
         &(\ln{X_n})_{xt}+X_{n-1}X_n^{-1}-X_nX_{n+1}^{-1}=\sum_{i=1}^K(q_{i,n}r_{i,n})_x),\\
         &q_{i,n,x}=(\ln{X_n})_{x}q_{i,n}+ q_{i,n-1},\qquad \qquad \quad i=1,2,\cdots, K ,\label{cn-2dTL-1}\\ 
        &r_{i,n,x}=-r_{i,n}(\ln{X_n})_{x}- r_{i,n+1},\qquad \qquad i=1,2,\cdots, K .\label{cn-2dTL-2}
\end{align}
On one hand, under the transformation 
\begin{equation*}
    x \rightarrow -x, \quad t\rightarrow -t, \quad X_n \rightarrow \exp(-X_n), \quad {q}_{i,n}\rightarrow-{q}_{i,n},
\end{equation*}
\eqref{cn-2dTL}-\eqref{cn-2dTL-2} becomes the nonlinear $2$DTL ECSC obtained by squared eigenfunction symmetries \cite{liu2008extended}
\begin{equation*}
    \begin{aligned}
         &(X_n)_{xt}=e^{X_n-X{n-1}}-e^{X_{n+1}-X_n}+\sum_{i=1}^K(q_{i,n}r_{i,n})_x,\\
         &q_{i,n,x}=-X_{n,x}q_{i,n}-q_{i,n-1},\quad i=1,2,\cdots, K,\\        &r_{i,n,x}=r_{i,n}X_{n,x}+r_{i,n+1},\qquad i=1,2,\cdots, K,
    \end{aligned}
\end{equation*}
On the other hand, under the transformation 
\begin{align*}
     x\rightarrow-x,\quad {X}_n=e^{-u_n},\quad   {q}_{i,n}=-\frac{1}{\sqrt{2}}w_{i,n}e^{-u_n},\quad
     {r}_{i,n}=\frac{1}{\sqrt{2}}v_{i,n+1}e^{u_n},
\end{align*}
\eqref{cn-2dTL}-\eqref{cn-2dTL-2} becomes the nonlinear $2$DTL ESCS obtained by source generation method \cite{wang20072d} 
    \begin{align}
        &u_{n,xt}+e^{u_n-u_{n-1}}-e^{u_{n+1}-u_n}=-\frac{1}{2}\sum_{i=1}^K(w_{i,n}v_{n+2,i}e^{u_{n+1}-u_n}-w_{n-1}v_{n+1}e^{u_n-u_{n-1}}),\label{wang-1}\\
        &\frac{w_{i,n,x}}{w_{i,n}}+\frac{w_{i,n-1}}{w_{i,n}}e^{u_n-u_{n-1}}=\frac{w_{i,n+1,x}}{w_{i,n+1}}+\frac{w_{i,n}}{w_{i,n+1}}e^{u_{n+1}-u_{n}},\label{wang-2}\\
        &\frac{v_{i,n,x}}{v_{i,n}}-\frac{v_{i,n+1}}{v_{i,n}}e^{u_n-u_{n-1}}=\frac{v_{i,n+1,x}}{v_{i,n+1}}-\frac{v_{n+2,i}}{v_{i,n+1}}e^{u_{n+1}-u_{n}}.\label{wang-3}
    \end{align}

It is noted that NC integrable systems can be treated in a unified way no matter what the reasons for noncommutativity are and results for NC integrable systems can produce the ones for the corresponding commutative integrable systems  which reaveals the advantage of dealing with NC integrable systems rather than commutative integrable systems.

\section{The matrix sG
ESCS and solutions}\label{sec4}
In \cite{li2008quasideterminant}, by imposing the $2$-periodic reductions on the non-Abelian $2$DTL equation, the matrix sG equation and its exact solutions were derived. 
In this section, we shall apply periodic and quasi-periodic reductions to the non-Abelian $2$DTL ESCS and its quasigrammian solutions to deduce a matrix sG ESCS as well as its exact solutions in terms of quasideterminants. 
 
Starting from the quasigrammian solutions $ {R}_n$, $ {q}_{i,n}$ and $ {r}_{i,n}$ given in Section \ref{3.2}, exact solutions  of the non-Abelian $2$DTL ESCS can be constructed as follows. 
For 
\begin{equation}\label{disper}
    (\theta_{i,n})_x=\theta_{i,n-1},\quad
    (\theta_{i,n})_t=\theta_{i,n+1},\quad
     (\rho_{i,n})_x=-\rho_{i,n+1},\quad
      (\rho_{i,n})_t=-\rho_{i,n-1},
\end{equation}
and $  {\Omega}_n$ is defined by \eqref{Omega-dispersion-1}. We choose the simplest non-trivial solutions  of \eqref{disper}
\begin{equation*}
    \theta_{k,n}=B_kb_k^{-n}\exp\left(b_kx+\dfrac{1}{b_k}t\right),\qquad \rho_{j,n}=A_ja_j^{n}\exp\left(-a_jx-\dfrac{1}{a_j}t\right),
\end{equation*}
where $A_j$ and $B_k$ are $d\times d$ matrices and then we obtain
\begin{equation*}
    {\Omega}_n(\theta_{k,n},\rho_{j,n})=\left(\dfrac{a_j}{b_k}\right)^n \left(\delta_{j,k}C_{j,k}(t)\left(\dfrac{b_k}{a_k}\right)^n+\dfrac{A_k^tB_ka_k}{b_k-a_k}\exp\left((b_k-a_k)x+\left(\dfrac{1}{b_k}-\dfrac{1}{a_k}\right)t\right)\right).
\end{equation*}
The choice of $\delta_{j,k}C_{j,k}(t)$ is needed to effect the periodic
reduction. Now using the invariance of a quasi-determinant to scaling of its rows and
columns \cite{gelfand2005quasideterminants}, we get
\begin{equation*}
   \begin{aligned}
     {R}_n&=-
      \left|\begin{matrix}
      \left(\delta_{j,k}C_{j,k}(t)\left(\dfrac{b_k}{a_j}\right)^n+\dfrac{A_j^tB_ka_j}{b_k-a_j}\exp\left(\xi_k+\hat{\xi_j}\right)\right) &   \left( A_j\exp\left(\hat{\xi_j}\right)\right)^{\dagger}\\
         \left(B_k\exp\left(\xi_k\right)  \right)    &     \boxed{-I}  
       \end{matrix}\right|,\\
      {q}_{i,n}&=\left|\begin{matrix}
      \left({a_j}^n \left(\delta_{j,k}C_{j,k}(t)\left(\dfrac{b_k}{a_j}\right)^n+\dfrac{A_j^tB_ka_j}{b_k-a_j}\exp\left(\xi_k+\hat{\xi_j}\right)\right) \right)  &     e_i^{\dagger}  \\
      \left( B_k\exp\left(\xi_k\right) \right)         &     \boxed{0}  
       \end{matrix}\right|\beta_i(t),   \\
    {r}_{i,n}&=\eta_i(t)
       \left|\begin{matrix}
       \left({b_k}^{-n} \left(\delta_{j,k}C_{j,k}(t)\left(\dfrac{b_k}{a_j}\right)^n+\dfrac{A_j^tB_k}{b_k(b_k-a_j)}\exp\left(\xi_k+\hat{\xi_j}\right)\right)\right)  &  \left( A_j\exp\left(\hat{\xi_j}\right)\right)^{\dagger} \\
                e_i                     &    \boxed{0}  
       \end{matrix}\right|,
    \end{aligned}
\end{equation*}
where $\xi_k=b_kx+\frac{1}{b_k}t$ and $\hat{\xi_j}=-a_jx-\frac{1}{a_j}t$.
If we further set $a_k=-b_k=\lambda_k$ with $\lambda_k$ being real scalars, which means that $(\frac{b_1}{a_1})^2=\cdots=(\frac{b_N}{a_N})^2=1$, we find that $ {R}_n$ and $ {q}_{i,n} {r}_{i,n+1}$ exhibit 2-periodicity with ${R}_{n+2}={R}_n,~{q}_{i,n+2} {r}_{i,n+3}={q}_{i,n} {r}_{i,n+1}$. 
However, $ {q}_{i,n}$ and $ {r}_{i,n}$ are quasi-periodic with
\begin{equation*}
    \begin{aligned} & {q}_{i,0}=\lambda_i^2 {q}_{i,2}=\cdots=\lambda_i^{2n} {q}_{i,2n},\qquad
     {q}_{i,1}=\lambda_i^2 {q}_{i,3}=\cdots=\lambda_i^{2n} {q}_{i,2n+1},\\
    & {r}_{i,0}=\lambda_i^{-2} {r}_{i,2}=\cdots=\lambda_i^{-2n} {r}_{i,2n},\quad
     {r}_{i,1}=\lambda_i^{-2} {r}_{i,3}=\cdots=\lambda_i^{-2n} {r}_{i,2n+1}.
    \end{aligned}
\end{equation*}
Under the above assumptions, the non-Abelian $2$DTL ESCS \eqref{2dTLSCS}-\eqref{2dTLSCS2} transforms into 
\begin{align}\label{nc-2period-sG}
   \begin{split}
       &(X_{0,x}X_0^{-1})_t+X_{1}X_0^{-1}- X_0X_{1}^{-1}=\sum_{i=1}^K(q_{i,0}r_{i,1}-q_{i,1}r_{i,2}),\\
       &(X_{1,x}X_1^{-1})_t+X_{0}X_1^{-1}-X_1X_{0}^{-1}=\sum_{i=1}^K(q_{i,1}r_{i,2}-q_{i,0}r_{i,1}),\\   
     & q_{i,0,x}=X_{0,x}X_0^{-1}q_0+{\lambda_i}^2q_{i,1}\qquad
       q_{i,1,x}=X_{1,x}X_1^{-1}q_{i,1}+q_{i,0},\\
       &r_{i,1,x}=-r_{i,1}X_{1,x}X_1^{-1}-r_{i,2},\qquad
       r_{i,2,x}=-r_{i,2}X_{0,x}X_0^{-1}-{\lambda_i}^2r_{i,1}
    \end{split} 
\end{align}
with solutions given by
\begin{equation*}
   \begin{aligned}
     {R}_0&=-
      \left|\begin{matrix}
       \left(\delta_{j,k}C_{j,k}(t)-\dfrac{A_j^tB_k\lambda_j}{\lambda_j+\lambda_k}\exp\left(\zeta_j+\zeta_k\right)\right) &    \left(A_j\exp\left(\zeta_j\right)\right)^{\dagger}\\
         \left(B_k\exp\left(\zeta_k\right) \right)     &     \boxed{-I}  
       \end{matrix}\right|,\\
       {R}_1&=-
      \left|\begin{matrix}
      \left( -\delta_{j,k}C_{j,k}(t)-\dfrac{A_j^tB_k\lambda_k}{\lambda_j+\lambda_k}\exp\left(\zeta_j+\zeta_k\right)\right) &    \left(A_j\exp\left(\zeta_j\right)\right)^{\dagger}\\
        \left( B_k\exp\left(\zeta_k\right) \right)     &     \boxed{-I}  
       \end{matrix}\right|,\\
        {q}_{i,0}&=\left|\begin{matrix}
      \left(\delta_{j,k}C_{j,k}(t)-\dfrac{A_j^tB_k\lambda_k}{\lambda_j+\lambda_k}\exp\left(\zeta_j+\zeta_k\right) \right)  &     e_i^{\dagger}  \\
      \left( B_k\exp\left(\zeta_k\right) \right)         &     \boxed{0}  
       \end{matrix}\right|\beta_i, \\
         {q}_{i,1}&=\left|\begin{matrix}
       \left( {\lambda_j} \left(-\delta_{j,k}C_{j,k}(t)-\dfrac{A_j^tB_k\lambda_k}{\lambda_j+\lambda_k}\exp\left(\zeta_j+\zeta_k\right)\right) \right)  &     e_i^{\dagger}  \\
      \left( B_k\exp\left(\zeta_k\right)  \right)        &     \boxed{0}  
       \end{matrix}\right|\beta_i, \\
    {r}_{i,1}&=\eta_i
       \left|\begin{matrix}
       \left(-\dfrac{1}{\lambda_k} \left(-\delta_{j,k}C_{j,k}(t)+\dfrac{A_j^tB_k}{\lambda_k(\lambda_j+\lambda_k)}\exp\left(\zeta_j+\zeta_k\right)\right) \right)  &  \left( A_j\exp\left(\zeta_j\right)\right)^{\dagger}\\
                e_i                     &    \boxed{0}  
       \end{matrix}\right|,\\
    {r}_{i,2}&=\eta_i
       \left|\begin{matrix}
       \left(\dfrac{1}{{\lambda_k}^2} \left(\delta_{j,k}C_{j,k}(t)+\dfrac{A_j^tB_k}{\lambda_k(\lambda_j+\lambda_k)}\exp\left(\zeta_j+\zeta_k\right)\right)\right)   &   \left(A_j\exp\left(\zeta_j\right)\right)^{\dagger} \\
                e_i                     &    \boxed{0}  
       \end{matrix}\right|,
    \end{aligned}
\end{equation*}
where $\zeta_j=-\lambda_jx-\frac{
1
}{\lambda_j}t$. 

From now on, we will assume that $A_j=I$ are real and $B_k$ is purely imaginary written as $\mathrm{i}r_kP_k$, where $r_k$ are real scalars. In this case, we have
\begin{equation*}
    \begin{aligned}
        {R}_0= {R}_1^{\ast},\quad q_i= {q}_{i,0}=\lambda_i {q}_{i,1}^{\ast},\quad r_i= {r}_{i,1}=\lambda_i^{-1} {r}_{i,2}^{\ast},
    \end{aligned}
\end{equation*}
where $^*$ means complex conjugate. By taking $X=R_0$,  \eqref{nc-2period-sG} are reduced to
\begin{equation}\label{matrix}
   \begin{aligned}&(X_{x}X^{-1})_t+X^{\ast}X^{-1}- X{X^{\ast}}^{-1}=\sum_{i=1}^K(q_{i}r_{i}-q_{i}^{\ast}r_{i}^{\ast}),\\ 
     & q_{i,x}=X_{x}X^{-1}q_i+{\lambda_i}q_{i}^{\ast},\\
       &r_{i,x}=-r_{i}X_{x}^{\ast}{X^{\ast}}^{-1}-\lambda_i r_{i}^{\ast}
   \end{aligned} 
\end{equation}
which we call a matrix sG ESCS henceforth. Actually, in the commutative case, by taking   
\begin{align*}
u=-2\mathrm{i}\log(X),\qquad
    q_{i}=\lambda_i(\phi_{i}-\mathrm{i}\psi_{i}),\qquad
     r_{i}=\mathrm{i}\phi_{i}+\psi_{i},\qquad 
\end{align*}
we have from the matrix sG ESCS \eqref{matrix} that


\begin{equation}\label{sGSCS}
   \begin{aligned}
      u_{xt}&=4\sin u+2\sum_{i=1}^K(\phi_{i}^2
                   +\psi_{i}^2)_x,\\
      \phi_{i,x}&=\frac{1}{2}u_x\psi_i+\lambda_i\phi_i,\\
      \psi_{i,x}&=-\frac{1}{2}u_x\phi_i-\lambda_i\psi_i,
  \end{aligned}
\end{equation}
which is nothing but the sG ESCS.
\begin{remark}
It is worthy pointing out that \eqref{sGSCS} is equivalent to sG ESCS studied in \cite{zhang2003n} 
\begin{equation*}
   \begin{aligned}
       u_{xt}&=\sin u+2\sum_{i=1}^K(\phi_{i}^2
                   +\psi_{i}^2)_x,\\
      \phi_{i,x}&=\frac{1}{2}u_x\psi_i-\lambda_i\phi_i,\\
      \psi_{i,x}&=-\frac{1}{2}u_x\phi_i+\lambda_i\psi_i
   \end{aligned}
\end{equation*}
 under the scaling transformations
\begin{equation*}
    x\rightarrow x,\quad t\rightarrow \frac{1}{4}t,\quad
    \phi_{i}\rightarrow 2\phi_{i},\quad \psi_{i}\rightarrow 2\psi_{i},\quad
    \lambda_i\rightarrow -\lambda_i.
\end{equation*}
\end{remark}

\section{Conclusions and discussions}\label{conclusion}
In this paper, we succeed in constructing and solving the non-Abelian $2$DTL ESCS by generalizing the source generation procedure to NC integrable systems. Two types of quasideterminant solutions for the non-Abelian $2$DTL ESCS are presented. In the case that $C_i(t)$ is independent of $t$, the non-Abelian $2$DTL ESCS reduces to the non-Abelian $2$DTL equation identically. By considering periodic and quasi-periodic reductions on the non-Abelian $2$DTL ESCS and its quasigrammian solutions, a matrix sG ESCS is proposed for the first time whose exact solutions in terms of quasideterminants are presented as well. As mentioned before, sources in SESCSs can result in solitary waves moving with nonconstant velocity or cause a great
variety of dynamics of soliton solutions, it is interesting to study dynamics of soliton solutions for NC versions of SESCSs.

\section*{Acknowledgement}
This work was supported by the National Natural Science
Foundation of China (Grants Nos. 11971322 and 12171475).

\appendix
 \section{Proof of the Proposition \ref{proposition-wron}} \label{sec:appendix-A}
Given 
 $Q_n$, $q_{i,n}$ and $r_{i,n}$ as defined in Section \ref{sub2.1}, we have
\begin{align*}  
     {Q}_{n,x}=&\left(
       \left|\begin{matrix}
         {\Theta}_{n-1}      &       \boxed{0}\\
       \tilde{  {\Theta}}_n    &        e_N   
       \end{matrix}\right|
     + \left|\begin{matrix}
         {\Theta}_{n}      &       \boxed{0}\\
       \tilde{  {\Theta}}_n    &       e_1   
       \end{matrix}\right|
     \left|\begin{matrix}
         {\Theta}_{n}      &       \boxed{0}\\
       \tilde{  {\Theta}}_n    &       e_N   
       \end{matrix}\right|  
       \right),\\
     ( {Q}_{n,x} {Q}_n^{-1})_t=&
    \left(\left|\begin{matrix}
            {\Theta}_n+H_{n-1}    &       \boxed{0}\\
          \tilde{  {\Theta}}_n      &        e_1   
          \end{matrix}\right|     
     +  \sum_{i=1}^N  \left|\begin{matrix}
                       {\Theta}_{n-1}        &       \boxed{0}\\
                     \tilde{  {\Theta}}_n      &        e_{i}   
                     \end{matrix}\right|
                     \left|\begin{matrix}
                       {\Theta}_{n+i+1}+H_{n+i}    &    \boxed{0}\\
                     \tilde{  {\Theta}}_n           &     e_{N}  
                     \end{matrix}\right|
   \right)\left|\begin{matrix}
         {\Theta}_n            &       \boxed{0}\\
       \tilde{  {\Theta}}_n      &        e_N   
       \end{matrix}\right|^{-1}\\      
   & -\left|\begin{matrix}
         {\Theta}_{n-1}        &       \boxed{0}\\
       \tilde{  {\Theta}}_n       &        e_N   
       \end{matrix}\right|
       \left|\begin{matrix}
         {\Theta}_{n}          &       \boxed{0}\\
       \tilde{  {\Theta}}_n       &        e_N   
       \end{matrix}\right|
         \left(\left|\begin{matrix}
           {\Theta}_{n+1}+H_{n}    &       \boxed{0}\\
         \tilde{  {\Theta}}_n         &        e_N   
         \end{matrix}\right|
   +  \sum_{i=1}^N   \left|\begin{matrix}
                       {\Theta}_{n}          &       \boxed{0}\\
                     \tilde{  {\Theta}}_n       &        e_{i}   
                     \end{matrix}\right|
                     \left|\begin{matrix}
                       {\Theta}_{n+i+1}    &    \boxed{0}\\
                     \tilde{  {\Theta}}_n             &     e_{N}  
                     \end{matrix}\right|\right.\\
   & +\left.\sum_{i=1}^N   \left|\begin{matrix}
                       {\Theta}_{n}          &       \boxed{0}\\
                     \tilde{  {\Theta}}_n       &        e_{i}   
                     \end{matrix}\right|
                \left|\begin{matrix}
                      H_{n+i}    &    \boxed{0}\\
                     \tilde{  {\Theta}}_n             &     e_{N}  
                     \end{matrix}\right|
   \right)
   \left|\begin{matrix}
         {\Theta}_n            &       \boxed{0}\\
       \tilde{  {\Theta}}_n       &        e_N   
       \end{matrix}\right|^{-1}
   +  \left|\begin{matrix}
                       {\Theta}_{n+1}+H_{n}    &    \boxed{0}\\
                     \tilde{  {\Theta}}_n        &     e_{1}  
                     \end{matrix}\right|\\
   &+  \sum_{i=1}^N  \left|\begin{matrix}
                       {\Theta}_{n}          &       \boxed{0}\\
                     \tilde{  {\Theta}}_n       &        e_{i}   
                     \end{matrix}\right|
                     \left|\begin{matrix}
                       {\Theta}_{n+i+1}+H_{n+i}    &    \boxed{0}\\
                     \tilde{  {\Theta}}_n             &     e_{1}  
                     \end{matrix}\right|\\
    =& {Q}_n {Q}_{n+1}^{-1}- {Q}_{n-1} {Q}_{n}^{-1}
      +\left|\begin{matrix}
          H_n               &     \boxed{0}\\
          \tilde{  {\Theta}}_n  &     e_1
        \end{matrix}\right|
      +  \sum_{i=1}^N 
        \left|\begin{matrix}
            {\Theta}_n               &     \boxed{0}\\
          \tilde{  {\Theta}}_n       &     e_i
        \end{matrix}\right|
        \left|\begin{matrix}
           H_{n+i}              &     \boxed{0}\\
          \tilde{  {\Theta}}_n      &     e_1
        \end{matrix}\right|\\
    &    -\left|\begin{matrix}
          H_{n-1}               &     \boxed{0}\\
          \tilde{  {\Theta}}_{n-1}  &     e_1
        \end{matrix}\right|
      -  \sum_{i=0}^{N-1} 
        \left|\begin{matrix}
            {\Theta}_{n-1}               &     \boxed{0}\\
          \tilde{  {\Theta}}_{n-1}       &     e_i
        \end{matrix}\right|
        \left|\begin{matrix}
           H_{n+i-1}                 &     \boxed{0}\\
          \tilde{  {\Theta}}_{n-1}       &     e_1
        \end{matrix}\right|.
\end{align*}
Notice the fact that $\beta(t)$ does not affect the proof concerning \eqref{2dTLSCS1}, we omit $\beta(t)$
\begin{align*}
     {q}_{i,n,x}=&\left|
    \begin{matrix}
          {\Theta}_n            &    \boxed{h_{i,n}}\\
        \tilde{  {\Theta}}_n   &    G_{i,n}
    \end{matrix}\right|_x
    =\left|
    \begin{matrix}
          {\Theta}_{n-1}            &    \boxed{h_{i,n-1}}\\
        \tilde{  {\Theta}}_n       &    G_{i,n}
    \end{matrix}\right|
    +  \sum_{j=1}^N 
    \left|
    \begin{matrix}
          {\Theta}_n            &   \boxed{0}\\
        \tilde{  {\Theta}}_n   &    e_j
    \end{matrix}\right| 
     \left|
  \begin{matrix}
          {\Theta}_{n-1+j}      &   \boxed{h_{i,n-1+j}}\\
        \tilde{  {\Theta}}_n   &    G_{n,i}
    \end{matrix}\right|\\
    =&\left|\begin{matrix}
          {\Theta}_{n-1}            &    \boxed{h_{i,n-1}}\\
        \tilde{  {\Theta}}_n       &    G_{i,n}
    \end{matrix}\right|
    +\left|\begin{matrix}
          {\Theta}_{n}              &    \boxed{0}\\
        \tilde{  {\Theta}}_n       &    e_1
    \end{matrix}\right|
    \left|\begin{matrix}
          {\Theta}_{n}              &    \boxed{h_{i,n}}\\
        \tilde{  {\Theta}}_n       &    G_{i,n}
    \end{matrix}\right|
\end{align*}
\begin{align*}
     {Q}_{n,x} {Q}_n^{-1} {q}_{i,n}+ {q}_{i,n-1}
=&\left(\left|\begin{matrix}
          {\Theta}_{n-1}       &   \boxed{0}\\
        \tilde{  {\Theta}}_n   &    e_N
    \end{matrix}\right|
    \left|\begin{matrix}
          {\Theta}_{n}       &   \boxed{0}\\
        \tilde{  {\Theta}}_n   &    e_N
    \end{matrix}\right|^{-1}
    +\left|\begin{matrix}
          {\Theta}_{n}         &   \boxed{0}\\
        \tilde{  {\Theta}}_n   &    e_1
    \end{matrix}\right|\right)
    \left|\begin{matrix}
          {\Theta}_{n}              &    \boxed{h_{i,n}}\\
        \tilde{  {\Theta}}_n       &    G_{i,n}
    \end{matrix}\right|\\
    &
    +\left|\begin{matrix}
          {\Theta}_{n-1}              &    \boxed{h_{i,n-1}}\\
        \tilde{  {\Theta}}_{n-1}     &    G_{i,n-1}
    \end{matrix}\right|.
\end{align*}
By making use of the NC Jacobi identity for quasi-determinants,  
we can obtain
\begin{align*}
    \left|\begin{matrix}
          {\Theta}_{n-1}              &    \boxed{h_{i,n-1}}\\
        \tilde{  {\Theta}}_{n}        &    G_{i,n}
    \end{matrix}\right|
    -\left|\begin{matrix}
          {\Theta}_{n-1}       &   \boxed{0}\\
        \tilde{  {\Theta}}_n   &    e_N
    \end{matrix}\right|
    \left|\begin{matrix}
          {\Theta}_{n}       &   \boxed{0}\\
        \tilde{  {\Theta}}_n   &    e_N
    \end{matrix}\right|^{-1}
    \left|\begin{matrix}
          {\Theta}_{n}              &    \boxed{h_{i,n}}\\
        \tilde{  {\Theta}}_n        &    G_{i,n}
    \end{matrix}\right|=
    \left|
    \begin{matrix}
        \tilde{  {\Theta}}_n   &    e_N     &   G_{i,n}\\
          {\Theta}_n           &    0       &   h_{i,n}\\
          {\Theta}_{n-1}       &    0       &   \boxed{h_{i,n-1}}
    \end{matrix}\right|.
\end{align*}
Notice that
\begin{align*}
   \left|
    \begin{matrix}
        \tilde{  {\Theta}}_n   &    G_{i,n}               &   e_N   \\
          {\Theta}_n           &   \boxed{h_{i,n-1}}      &   0\\
          {\Theta}_{n-1}       &    h_{i,n}               &   0
    \end{matrix}\right| 
    =\left|
    \begin{matrix}
        \tilde{  {\Theta}}_{n-1}   &    G_{i,n-1}           &   0   \\
          {\Theta}_{n-1}           &   \boxed{h_{i,n-1}}  &   0  \\
          {\Theta}_{n+1}           &    h_{i,n+N}         &   1
    \end{matrix}\right| 
    =\left|
    \begin{matrix}
        \tilde{  {\Theta}}_{n-1}   &    G_{i,n-1}     \\
          {\Theta}_{n-1}           &   \boxed{h_{i,n-1}}
    \end{matrix}\right|,
\end{align*}
it is clear that \eqref{2dTLSCS1} holds true.
It is not difficult to prove \eqref{2dTLSCS2} by noticing
\begin{align}\label{identity}
    \left|\begin{matrix}
         e_i^{T}     &    \boxed{0}\\
        \tilde{  {\Theta}}_{n-1}  &    e_2
    \end{matrix}\right|
    =\left|\begin{matrix}
         e_i^{T}     &    \boxed{0}\\
        \tilde{  {\Theta}}_{n-1}  &    e_1
    \end{matrix}\right|
    \left|\begin{matrix}
           {\Theta}_{n}     &    \boxed{0}\\
        \tilde{  {\Theta}}_{n}  &    e_1
    \end{matrix}\right|
    +\left|\begin{matrix}
         e_i^{T}     &    \boxed{0}\\
        \tilde{  {\Theta}}_{n}  &    e_1
    \end{matrix}\right|,
\end{align} and 
\begin{align*}
     {r}_{i,n,x}=&
    \sum_{j=1}^N 
   \left|\begin{matrix}
         e_i^{I}     &    \boxed{0}\\
        \tilde{  {\Theta}}_{n-1}  &    e_j
    \end{matrix}\right|
    \left|\begin{matrix}
           {\Theta}_{n-2+j}     &    \boxed{0}\\
        \tilde{  {\Theta}}_{n-1}  &    e_1
    \end{matrix}\right|
    = \left|\begin{matrix}
         e_i^{t}     &    \boxed{0}\\
        \tilde{  {\Theta}}_{n-1}  &    e_1
    \end{matrix}\right|
     \left|\begin{matrix}
           {\Theta}_{n-1}     &    \boxed{0}\\
        \tilde{  {\Theta}}_{n-1}  &    e_1
    \end{matrix}\right|
    -\left|\begin{matrix}
         e_i^{T}     &    \boxed{0}\\
        \tilde{  {\Theta}}_{n-1}  &    e_2
    \end{matrix}\right|,
\end{align*}
\begin{align*}
 {r}_{i,n} {Q}_{n,x} {Q}_n^{-1}+  {r}_{i,n+1}
=&\left|\begin{matrix}
         e_i^{t}     &    \boxed{0}\\
        \tilde{  {\Theta}}_{n-1}  &    e_1
    \end{matrix}\right|
    \left(\left|\begin{matrix}
           {\Theta}_{n-1}     &    \boxed{0}\\
        \tilde{  {\Theta}}_{n}  &    e_N
    \end{matrix}\right|
    \left|\begin{matrix}
           {\Theta}_{n}     &    \boxed{0}\\
        \tilde{  {\Theta}}_{n}  &    e_N
    \end{matrix}\right|^{-1}
    +\left|\begin{matrix}
           {\Theta}_{n}     &    \boxed{0}\\
        \tilde{  {\Theta}}_{n}  &    e_1
    \end{matrix}\right|\right)
    + \left|\begin{matrix}
         e_i^{T}     &    \boxed{0}\\
        \tilde{  {\Theta}}_{n}  &    e_1
    \end{matrix}\right|.
\end{align*}


\section{Proof of the quasigrammian solutions}\label{appendixB}
By detailed calculations, we have the following derivative formulae for $R_n$, $q_{i,n}$ and $r_{i,n}$ as defined in Section \ref{3.2},
\begin{align*}
     {R}_n^{-1}=&\left|\begin{matrix}
         {\Omega}_{n-1}  &   P_n^{\dag}\\
            \Theta_n        &   \boxed{I}
    \end{matrix}\right|
    =I-\Theta_n  {\Omega}_{n-1}^{-1}P_n^{\dag},\\
    ( {R}_{n,x} {R}_n^{-1})_t=&I-\Theta_n {\Omega}_{n-1}^{-1}P_n^{\dag}+\Theta_{n-1} {\Omega}_{n-1}^{-1}P_{n-1}^{\dag}
    -\Theta_{n-1} {\Omega}_{n-1}^{-1}P_{n-1}^{\dag}\Theta_n {\Omega}_{n-1}^{-1}P_n^{\dag}\\
    &-I+\Theta_{n+1} {\Omega}_{n}^{-1}P_{n+1}^{\dag}+\Theta_n^{-1} {\Omega}_{n}^{-1}P_{n}^{\dag}+\Theta_{n} {\Omega}_{n}^{-1}P_{n}^{\dag}\Theta_{n+1} {\Omega}_{n}^{-1}P_{n+1}^{\dag}\\
    &+\sum_{i=1}^K(\Theta_n {\Omega}_{n}^{-1}c'_i {\Omega}_{n}^{-1}P_{n+1}^{\dag}-\Theta_{n-1} {\Omega}_{n-1}^{-1}c'_i {\Omega}_{n-1}^{-1}P_{n}^{\dag}),\\
     {R}_{n-1} {R}_n^{-1}=&I-\Theta_n {\Omega}_{n-1}^{-1}P_n^{\dag}+\Theta_{n-1} {\Omega}_{n-1}^{-1}P_{n-1}^{\dag}
    -\Theta_{n-1} {\Omega}_{n-1}^{-1}P_{n-1}^{\dag}\Theta_n {\Omega}_{n-1}^{-1}P_n^{\dag},\\
     {R}_n {R}_{n+1}^{-1}=&I-\Theta_{n+1} {\Omega}_{n}^{-1}P_{n+1}^{\dag}-\Theta_n^{-1} {\Omega}_{n}^{-1}P_{n}^{\dag}-\Theta_{n} {\Omega}_{n}^{-1}P_{n}^{\dag}\Theta_{n+1} {\Omega}_{n}^{-1}P_{n+1}^{\dag},\\
     {q}_{i,n,x}=&\left|\begin{matrix}
        {\Omega}_{n}    &     e_i\\
       \Theta_n            &    \boxed{0}
   \end{matrix}\right|_x=-(\Theta_n {\Omega}_{n}^{-1}e_i)_x=-\Theta_{n-1} {\Omega}_{n}^{-1}e_i+\Theta_n {\Omega}_{n}^{-1}P_{n+1}^{\dag}\Theta_n {\Omega}_{n}^{-1}e_i,\\
    {R}_{n,x} {R}_n^{-1} {q}_{i,n}=&\Theta_n {\Omega}_{n}^{-1}P_{n+1}^{\dag}\Theta_n {\Omega}_{n}^{-1}e_i-\Theta_{n-1} {\Omega}_{n-1}^{-1}P_{n}^{\dag}\Theta_n {\Omega}_{n}^{-1}e_i\\
   =&\Theta_n {\Omega}_{n}^{-1}P_{n+1}^{\dag}\Theta_n {\Omega}_{n}^{-1}e_i+\Theta_{n-1} {\Omega}_{n-1}^{-1}e_i-\Theta_{n-1} {\Omega}_{n}^{-1}e_i,\\
    {r}_{i,n,x}=&\left|\begin{matrix}
        {\Omega}_{n-1}    &     P_n^{\dag}\\
        e_i^{\dag}           &    \boxed{0}
   \end{matrix}\right|_x=-( e_i^{\dag} {\Omega}_{n-1}^{-1}P_n^{\dag})_x=e_i^{\dag} {\Omega}_{n-1}^{-1}P_{n+1}^{\dag}-e_i^{\dag} {\Omega}_{n-1}^{-1}P_n^{\dag}\Theta_{n-1} {\Omega}_{n-1}^{-1}P_n^{\dag},\\
    {r}_{i,n} {R}_{i,n,x} {R}_n^{-1}=&-e_i^{\dag} {\Omega}_{n-1}^{-1}P_n^{\dag}\Theta_{n} {\Omega}_{n}^{-1}P_{n+1}^{\dag}+e_i^{\dag} {\Omega}_{n-1}^{-1}P_n^{\dag}\Theta_{n-1} {\Omega}_{n-1}^{-1}P_n^{\dag}\\
   =&e_i^{\dag} {\Omega}_{n-1}^{-1}P_{n+1}^{\dag}-e_i^{\dag} {\Omega}_{n}^{-1}P_{n+1}^{\dag}+e_i^{\dag} {\Omega}_{n-1}^{-1}P_n^{\dag}\Theta_{n-1} {\Omega}_{n-1}^{-1}P_n^{\dag},
\end{align*}
which are sufficient to prove that $R_n$, $q_{i,n}$ and $r_{i,n}$ provide quasigrammian solutions of the non-Abelian $2$DTL ESCS. 

\normalem 

\end{document}